\documentstyle[12pt,openbib]{article}
\begin{document}
\title{\bf  Problem of Statistical Model in Deep Inelastic
Scattering Phenomenology }
\baselineskip .5cm
\author{Jishnu Dey\thanks{Supported by FAPESP of S\~ao Paulo, Brasil} and
Lauro Tomio \\
Instituto de F\'\i sica Te\'orica,
Universidade Estadual Paulista \\
01405-900, S\~ao Paulo, SP, Brasil \\
Mira Dey\thanks{on
leave from Department of Physics, Lady Brabourne College, Calcutta 700017
India and supported in part by DST (SP/S2/K04/92), Govt. of India} \\ ICTP,
P.O. Box 586, Trieste, Italy 34100
}
\baselineskip .7cm
\vspace{.5 cm}
\date{\today }
\maketitle
{\it Abstract} :

Recent Deep Inelastic data leads to an up-down quark asymmetry
of the nucleon sea.  Explanations of the flavour asymmetry and
the di-lepton production in proton-nucleus collisions call for a
temperature $T \approx 100$ MeV in a statistical model. This T
may be conjectured as being due to the Fulling-Davies-Unruh
effect. But it is not possible to fit the structure function
itself.
\newpage

There is lot of excitement because of availability of very
precise data on the structure functions of the proton and the
neutron (deuteron) by the NMC collaboration\cite{nmc}. They find
that the structure functions for proton and neutron violates the
Gottfried sum rule \cite{gott} implying that the sea is not
flavour symmetric. There is more d anti-quark in proton sea than
the u antiquark.

The Fermilab experiment E772 reported the measurements of the
yields of di-muons in a 800 GeV protons colliding with
isoscalar and neutron excess targets \cite{mc}. The asymmetry in
$\bar u(x)$ and $\bar d(x)$ distribution will have its mark in
the cross section. Any model describing the Gottfried defect
should also explain this Drell-Yan process.

The process $P+A \Rightarrow \mu ^{+}\mu ^{-}+X$ is dominated in
low $X_F$ region by a quark $q$ of a particular flavour
annihilating the antiquark $\bar q$ of same flavour. Hence the
cross-section is sensitive to the distribution of the anti-quark
in the target nucleus. In the E772 experiment the target nuclei
were isoscalars $^2H$, C and neutron rich W. The ratio of
cross-sections per nucleon $\sigma _A$ in proton collision with
a nucleus A, to that $\sigma _{iso}$, with an isoscalar target,
is given by

\begin{equation}
R_A \equiv {{\sigma _A \over \sigma _{iso}}\approx
1+{(N-Z)\over A}{{\bar d(x)-\bar u(x)}\over {\bar d(x)+\bar
u(x)}}}
\end{equation}

The differential cross section is given by

\begin{equation}
{ m^3 {d \sigma \over {dm \; dX_F}} = {{8\pi \alpha ^2}\over
9}{\tau \over {\sqrt {X^{2}_F+4\tau }}}{\sum_i e^{2}_{q_i}
[q_i(x_1)\bar q_i(x_2)+ (1\leftrightarrow 2)]}}
\end{equation}

where $\tau = m^2/s$, $m^2=x_1x_2s$ is the di-muon invariant
mass square. $x_1$ and $x_2$ are the momentum fraction for the
$q$ and the $\bar q$, $X_F= x_1-x_2$ the Feynman variable. $s= 2 \times
M \times E$, where $M$ is the mass of the proton and $E$ is the beam
energy.

We go back to NMC experiment. The difference in the structure
function for proton and neutron when integrated gives rise to

\begin{equation}
S_G\equiv \int _{0.004}^{0.8}{[F_2^n(x) - F_2^p(x)]
\frac{dx}{x}} = 0.227 \pm 0.007\; (stat)\pm 0.014\;(sys)
\label{eq:exp}
\end{equation}

It is claimed \cite{zol} that when extrapolated to $x = 0$ to $1$,
this integral may give values much less, about half of the 1/3
expected from symmetric sea ! The unpolarized structure
functions of proton and neutron in the quark-parton model are
respectively given by

\begin{equation}
F_2^{e,p}(x) = \frac{4}{9}[u^p(x) + \bar{u}^p(x)] +
\frac{1}{9}[d^p(x) + \bar{d}^p(x)],
\end{equation}

\begin{equation}
F_2^{e,n}(x) = \frac{4}{9}[u^n(x) + \bar{u}^n(x)] +
\frac{1}{9}[d^n(x) + \bar{d}^n(x)].
\label{eq:gott}
\end{equation}

Assuming $u^p = d^n = u$ and $d^p = u^n = d$, the
eqn.(\ref{eq:exp}) leads to

\begin{equation}
S_G=\frac{1}{3}+ \frac{2}{3}\int _o^1 \left[ \bar{u}(x)  -
\bar{d}(x)\right] dx
\end{equation}\\ so that

\begin{equation}
I_G \equiv \int _0^1{[\bar{u}(x) - \bar{d}(x)] dx} = - 0.140\pm 0.024
\label{eq:anti}
\end{equation}

The valence quark distributions in the proton is

\begin{equation}
\int _o^1{u_v(x) dx} \equiv  \int _o^1 \left[ u(x) - \bar{u}(x)\right] dx
 = 2,
\label{eq:nu}
\end{equation}\\ and

\begin{equation}
\int_o^1 {d_v(x) dx} \equiv  \int _o^1\left[ d(x) -
\bar{d}(x)\right] dx  = 1.
\label{eq:nd}
\end{equation}

It is evident from eqns. (1) and (\ref{eq:anti}) that there is a
consistency requirement for  the distribution function.

In the ref. \cite{mc} the data for the Drell-Yan ratio have been
compared with different model-calculations \cite{es}, \cite{ehq}
and \cite{kl}. Although all the models were consistent with NMC
data only the last model was found to be within the experimental
error bars for the Drell-Yan ratio. Later, Eichten, Hinchliffe
and Quigg \cite{EHQ} showed that their model is consistent with
NMC as well as E772 experiments. In this perspective let us
investigate what the statistical model predicts.

In this model the partons are described as a gas inside the
confining hadron at finite temperature. \ Although very
speculative, the model has been studied by many workers in the
field \cite{gas}, \cite{mac}.

How does one reconcile oneself with such a temperature?
We have pointed out earlier that this temperature could
possibly arise because of Fulling-Davies-Unruh effect (FDU)
\cite{FDU} : accelerating particles feel a hot vacuum
\cite{ddts}. The applicability of FDU to Dirac particles was
first treated by Soffel, M\"uller and Greiner \cite{sof}.  This
should also be of importance to hadron physics, since light
quarks encounter very rapid change of velocity at the confining
boundary of hadrons. The velocity of quarks is nearly equal to
that of light, even in the constituent quark model and at the
border of the confining region the quark must turn back sharply,
in order that the confinement paradigm, to which we subscribe,
should be valid.

But one has to be careful about the non-uniform
acceleration.  It is known that a uniformly accelerated detector
in the Minkowski vacuum feels a thermal bath characterized by a
T, proportional to its proper acceleration. For non-uniform
acceleration, or for example a sudden deceleration felt by a
lepton during a DIS process, - it is not at all clear that a
simple thermal bath picture is adequate. Even for uniform
acceleration there are problems of divergence in the excitation
rate for finite-time detectors
\cite{sva} ; only recently it has been shown that no
divergence appears provided the detectors are turned on and off
continuously as in  a more realistic picture  for modeling
physical detectors \cite{hig}.

An effective temperature in a hadron may affect the structure
functions and in particular may affect the difference effect in
neutrons and protons. \ We shall, in particular, follow the work
of Mac and Ugaz \cite{mac}. \ In their language, one could not
``ascribe to the effective temperature, for example, any deep
physical meaning (or permanent physical reality) concerning such
a complicated bound system as the nucleon". \ In our point of
view, the presence of some kind of an average temperature is
natural, once one admits that the fast moving quarks have large
average accelerations $a$.

 We looked at Gottfried sum rule, which addresses the difference
in the structure functions of the proton and the neutron and  simultaneously at
the related problem of the distribution
function in Drell-Yan process. The parameters of the statistical
model are the temperature T, chemical potentials for the u and d
and the radius of the nucleon. We intend to fit these parameters
to get the phenomenology right.

 We cannot hope to fit the proton structure function itself,
since we are using a model where quarks are bound, and as
pointed out by Reya \cite{reya} all bound state approaches to
DIS have problem since the scale of bound state problem is 100
$MeV$. This is well known at large $x$, close to 1 where bound
state structure functions do not go to zero. In a recent paper
Donnachie and Landshoff \cite{don}  analyzed the problem from
the phenomenological point of view. They point out that the
variation of the structure function $\nu W_2$ with $Q^2$ at
small $Q^2$ cannot be described by perturbative QCD : it is
unsafe to use any perturbative evolution equation until $Q^2$ is
at least so large that $\nu W_2$  has fully {\it recovered} from
its need to vanish at $Q^2 = 0$. The latest NMC data on the
structure function at small $x$ \cite{nmc1} is contrary to all
earlier expectation and theoretical fits ! It is found to
increase at small $x$. This is incorporated in the ref.
\cite{don} along with the real photon data, which makes the fit
very attractive. It may be that some bound state model can meet
this phenomenological model halfway, when it has {\it recovered}
from its boundedness. It is missing in our model.

As  mentioned above, a defect of the model of Mac and Ugaz or
any bound state model for partons is that for $x$ = 1, the
structure functions do not go to zero. It appears the model
tries to ameliorate this problem by choosing a radius which is
very large, about 2 fm. This is unsatisfactory, but the point of
current interest is the antiparticle distributions and this is
only substantial for small $x$. So we expect the model that we
have adopted reproduces the essential physics of small $x$ DIS.

We hope that since the Gottfried sum rule refers to the
difference between the proton and the neutron, the large $x$ part
cancels out. The structure functions show asymmetry in spin and
isospin only at small $x$.  This was pointed out in the papers by
Carlitz and Kaur \cite{car} and Kaur
\cite{kaur}.

We start with the mean number of quarks with two polarizations
and momentum within {\bf $ p$} to {\bf $p+dp$}:

\begin{equation}
q_i({\bf p}) = \frac {6 V}{(2 \pi )^3} \left[ 1 + \exp {\frac
{\epsilon - \mu _i} {T}}\right]^{-1},
\label{eq:fermi}
\end{equation}
where $i$ is the flavour label, $\mu _i$ is the chemical
potential for the respective quark, V is the volume and
$\epsilon $ the corresponding energy.  The quark distribution in
the infinite momentum frame is given by Mac and Ugaz \cite{mac} to be

\begin{equation}
q_i(x) = \frac {6 V}{(2 \pi )^2} M^2 T x  ln\left[ 1 +
\exp {\frac {\mu _i - Mx/2}{T}}\right],
\label{eq:mac}
\end{equation}
where now  $q_i(x) dx$  is the probability of finding a quark
carrying the momentum fraction between $x$ and  $x + dx$ of the
total nucleon momentum.

Fixing M = 938 $MeV$ we find the parameter set $T = 103\; MeV$, $\mu _{u} = 148
\; MeV$, $\mu _{d} =
83.4 \; MeV$ and $R = 1.28 \; fm$  giving  eqns.
(\ref{eq:nu}, \ref{eq:nd}) as $$
\begin{array}{lll}
S_G & = &0.22\\
I_G & = &0.14
\end{array}
$$
They compare very well with the experimental numbers (eqns.
\ref{eq:exp} and \ref{eq:anti}).

In Figs. 1 and 2 we plot the Drell-Yan ratio and the
differential cross section (eqns. 1-2). The fit seems to be good. Regarding
$F_2$ itself
the situation is hopeless. Our  $F_2$ has a peak of about 0.7 -
0.8 and then goes down, whereas the recent data show a flat
structure. In \cite{mac} they have given formulae for using four
particle gluon graphs with an extra parameter K, which measures
the gluon coupling. The value of all quantities including $F_2$,
now involves a tedious integral over the gluon variable $y$.
But performing extensive time-consuming searches we found that
we cannot improve the fits even with the extra available
parameter K. We should also mention that the lower temperature given in refs.
[9-10] is inadequate to explain Gottfried defect and Drell-Yan processes.  So
we conclude that a simple statistical model can
give a rough description of the antiparticle cloud, but a better
model is certainly necessary to understand all aspects of deep
inelastic scattering.

  One of us (MD) is grateful to Professor Abdus Salam,
the IAEA and the UNESCO for hospitality at ICTP. It is a
pleasure to thank K. Sridhar, Paulo Leal Ferreira, George
Matsas, Clovis Peres, Tobias Frederico and Marcelo Schiffer for
valuable criticism.
\

\vspace {.5 cm}

{\bf Figure caption}

Fig. 1 : Drell-Yan Ratio $R_A$ (eqn. 1) vs. x. Experimental data are from
\cite{mc}.

Fig. 2 : Drell-Yan differential cross-section (eqn. 2) vs $X_F$ . Experimental
data are from \cite{mc}.

\vfill

\vskip .5cm
\begin{thebibliography}{99}


\bibitem{nmc} NMC Collaboration, P. Amaudruz et al., Phys. Rev.
Lett., 66 (1991) 2712

\bibitem{gott} K. Gottfried, Phys. Rev. Lett. 18 (1967) 1154.

\bibitem{mc} E 772 collaboration, P.L. McGaughey  et al, Phys.
Rev. Lett., 69 (1992) 1726

\bibitem{zol} V. R. Zoller, Z. Phys. C 54 (1992) 425, Phys. Lett.
B 279 (1992) 145.
\bibitem{es} S.D. Ellis and W.J. Stirling, Phys. Lett. B 256
(1991) 258

\bibitem{ehq} E. Eichten, I. Hinchliffe and C. Quigg, Phys. Rev.
D 45 (1992) 2269

\bibitem{kl} S. Kumano and J.T. Londergan, Phys. Rev. D 44 (1991) 717

\bibitem{EHQ} E. J. Eichten, I. Hinchliffe and C. Quigg, Phys.
Rev. D 47 (1993) 747

\bibitem{gas}C. Angelini and R. Pazzi, Phys. Lett. B 133 (1982)
343 ; J. Chela-Flores, E. Ugaz, Lett. Nuovo Cim. 38 (1983) 410 ;
\ T. Kawabe, Lett. Nuovo Cim. 44 (1985) 150 ; J. Cleymans and
R. L. Thews, Z. Phys. C 37 (1988) 315 ; R. P. Bickerstaff and
J.T. Londergan, Phys. Rev. D 42 (1990) 90 ; K. Ganesamurthy, V.
Devanathan and M. Rajasekaran, Z. Phys. C 52 (1991) 589

\bibitem{mac} E. Mac and E. Ugaz, Z.  Phys. C 43 (1989) 655; \

\bibitem{FDU} S. A. Fulling, Phys. Rev. D7 (1973) 2850 ; P.
Davies, J.  Phys. A8 (1975) 609 ; W. Unruh, Phys.  Rev. D 14
(1976) 870 ; S.  A. Fulling, {\it{Aspects of Quantum Field
Theory in Curved Space Time}}, (Cambridge Univ. Press 1989).

\bibitem{ddts} J. Dey, M. Dey, L. Tomio and M. Schiffer, Phys.
Lett. A 172 (1993) 203.

\bibitem{sof} M. Soffel, B. M\"uller and W. Greiner, Phys. Rev.
D 22 (1980) 1935.

\bibitem{sva} B. F. Svaiter and N. F. Svaiter, Phys. Rev. D 46,
(1992) 5267.

\bibitem{hig} A. Higuchi, G. E. A. Matsas and C. B. Peres,
Phys. Rev. D (in press), IFT-P034/93 ; G. E. A. Matsas and C. B.
Peres, IFT-P039/93 to be published.

\bibitem{reya} E. Reya, {\it The Spin Structure of the Nucleon},
Plenary talk presented at the Workshop on QCD - 20 Years Later
(World Scientific, Singapore, 1993). Dortmund preprint DO-TH
92-17, p. 28.

\bibitem{nmc1} NMC collaboration, Phys. Lett. B 295 (1992) 159

\bibitem{don} A. Donnachie and P.V. Landshoff, {\it Proton
Structure Function at Small $Q^2$}, Preprint, M/C-TH
93/11, DAMTP 93-23

\bibitem{car} R. Carlitz and J. Kaur, Phys. Rev. Lett. 37 (1977) 673.

\bibitem{kaur} Jasjit Kaur, Nucl. Phys. B128 (1977) 219.



\end{thebibliography}
\end{document}